\documentclass{revtex4-2}%
\usepackage[latin9]{inputenc}
\usepackage{amsmath}
\usepackage{amssymb}
\usepackage{graphicx}
\usepackage[unicode=true,  bookmarks=false,  breaklinks=false,pdfborder={0 0
1},backref=section,colorlinks=false]{hyperref}
\usepackage{amsfonts}
\usepackage{subfigure}
\usepackage{cleveref}
\usepackage{listings}
\usepackage{xcolor}
\usepackage{array}
\usepackage{url}%
\hypersetup{
 colorlinks,linkcolor=blue,citecolor=blue,urlcolor=blue}
\makeatletter
\@ifundefined{textcolor}{}
{
\definecolor{BLACK}{gray}{0}
 \definecolor{WHITE}{gray}{1}
 \definecolor{RED}{rgb}{1,0,0}
 \definecolor{GREEN}{rgb}{0,1,0}
 \definecolor{BLUE}{rgb}{0,0,1}
 \definecolor{CYAN}{cmyk}{1,0,0,0}
 \definecolor{MAGENTA}{cmyk}{0,1,0,0}
 \definecolor{YELLOW}{cmyk}{0,0,1,0}
}

\makeatother
\begin{document}
\preprint{CTP-SCU/2021003}
\title{Scalarized Einstein-Maxwell-scalar Black Holes in Anti-de Sitter Spacetime}
\author{Guangzhou Guo}
\email{guangzhouguo@stu.scu.edu.cn}
\author{Peng Wang}
\email{pengw@scu.edu.cn}
\author{Houwen Wu}
\email{iverwu@scu.edu.cn}
\author{Haitang Yang}
\email{hyanga@scu.edu.cn}
\affiliation{Center for Theoretical Physics, College of Physics, Sichuan University,
Chengdu, 610064, China}

\begin{abstract}
In this paper, we study spontaneous scalarization of asymptotically anti-de
Sitter charged black holes in the Einstein-Maxwell-scalar model with a
non-minimal coupling between the scalar and Maxwell fields. In this model,
Reissner-Nordstr\"{o}m-AdS (RNAdS) black holes are scalar-free black hole
solutions, and may induce scalarized black holes due to the presence of a
tachyonic instability of the scalar field near the event horizon. For RNAdS
and scalarized black hole solutions, we investigate the domain of existence,
perturbative stability against spherical perturbations and phase structure. In
a micro-canonical ensemble, scalarized solutions are always thermodynamically
preferred over RNAdS black holes. However, the system has much rich phase
structure and phase transitions in a canonical ensemble. In particular, we
report a RNAdS BH/scalarized BH/RNAdS BH reentrant phase transition, which is
composed of a zeroth-order phase transition and a second-order one.

\end{abstract}
\maketitle
\tableofcontents

{}

\section{Introduction}

The no-hair theorem states that a black hole can be uniquely determined via
three parameters, its mass, electric charge and angular momentum
\cite{Israel:1967wq,Carter:1971zc,Ruffini:1971bza}. Although this theorem
holds true in the Einstein-Maxwell field theory, it suffers from challenges
due to the existence of hairy black holes possessing extra macroscopic degrees
of freedom in other theories. In fact, various black hole solutions, e.g.,
hairy black holes in the Einstein-Yang-Mills theory
\cite{Volkov:1989fi,Bizon:1990sr,Greene:1992fw}, black holes with Skyrme hairs
\cite{Luckock:1986tr,Droz:1991cx} and black holes with dilaton hairs
\cite{Kanti:1995vq}, have been discovered as counter-examples to the no-hair
theorem. For a review, see \cite{Herdeiro:2015waa}.

Spontaneous scalarization typically occurs in models with non-minimal coupling
terms of scalar fields, which can source the scalar fields to destabilize
scalar-free black hole solutions and form scalarized hairy black holes. This
phenomenon was first studied for neutron stars in scalar-tensor models
\cite{Damour:1993hw} by coupling scalar fields to the Ricci curvature. It
demonstrated that there is a coexistence region for scalar-free and scalarized
neutron star solutions, where the scalarized ones can be energetically
preferred. Later, it was found that there also exists spontaneous
scalarization of black holes in scalar-tensor models, provided that black
holes are surrounded by non-conformally invariant matter
\cite{Cardoso:2013opa,Cardoso:2013fwa}.

Recently, the phenomenon of spontaneous scalarization has been studied in the
extended Scalar-Tensor-Gauss-Bonnet (eSTGB) gravity
\cite{Doneva:2017bvd,Silva:2017uqg,Antoniou:2017acq,Cunha:2019dwb,Herdeiro:2020wei,Berti:2020kgk}%
. In particular, asymptotically anti-de Sitter (AdS) scalarized black holes
have been studied in a scalar-tensor model with non-minimally coupling the
scaler field to the Ricci scalar and the Gauss-Bonnet term, as well as
applications to holographic phase transitions \cite{Brihaye:2019dck}. In eSTGB
models, the scalar field is non-minimally coupled to the Gauss-Bonnet
curvature correction of the gravitational sector, which leads to numerical
challenges for solving the evolution equations due to non-linear curvature
terms. To better understand dynamical evolution into scalarized black holes, a
similar, but technically simpler, type of models, i.e.,
Einstein-Maxwell-scalar (EMS) models with non-minimal couplings between the
scalar and Maxwell fields, have been put forward in \cite{Herdeiro:2018wub},
where fully non-linear numerical evolutions of spontaneous scalarization were
presented. Subsequently, further studies of spontaneous scalarization in the
EMS models were discussed in the context of various non-minimal coupling
functions \cite{Fernandes:2019rez,Blazquez-Salcedo:2020nhs}, dyons including
magnetic charges \cite{Astefanesei:2019pfq}, axionic-type couplings
\cite{Fernandes:2019kmh}, massive and self-interacting scalar fields
\cite{Zou:2019bpt,Fernandes:2020gay}, horizonless reflecting stars
\cite{Peng:2019cmm}, stability analysis of scalarized black holes
\cite{Myung:2018vug,Myung:2019oua,Zou:2020zxq,Myung:2020etf,Mai:2020sac}, higher
dimensional scalar-tensor models \cite{Astefanesei:2020qxk}, quasinormal modes
of scalarized black holes \cite{Myung:2018jvi,Blazquez-Salcedo:2020jee}, two
U(1) fields \cite{Myung:2020dqt}, quasi-topological electromagnetism
\cite{Myung:2020ctt}, topology and spacetime structure influences
\cite{Guo:2020zqm} and the Einstein-Born-Infeld-scalar theory
\cite{Wang:2020ohb}. Besides the above asymptotically flat scalarized black
holes, spontaneous scalarization was also discussed in the EMS model with a
positive cosmological constant \cite{Brihaye:2019gla}. Additionally,
spontaneous vectorization of electrically charged black holes was also
proposed \cite{Oliveira:2020dru},  analytic treatments were applied to
study spontaneous scalarization in the EMS models
\cite{Konoplya:2019goy,Hod:2020ljo,Hod:2020ius,Hod:2020cal}, and an infinite family of exact topological charged hairy black hole solutions were constructed in the EMS gravity system \cite{Mahapatra:2020wym}. 

Studying thermodynamics of the EMS models not only provides evidence for
endpoints of the dynamical evolution of unstable scalar-free black holes, but
also is interesting per se. Understanding the statistical mechanics of black
holes has been a subject of intensive study for several decades. In the
pioneering work \cite{Hawking:1974sw,Bekenstein:1972tm,Bekenstein:1973ur},
Hawking and Bekenstein found that black holes possess the temperature and the
entropy. However, asymptotically flat black holes are often thermally unstable
since they have negative specific heat. To make black holes thermally stable,
appropriate boundary conditions need to be imposed, e.g., putting the black
holes in AdS space. Asymptotically AdS black holes become thermally stable
since the AdS boundary acts as a reflecting wall. The thermodynamic properties
of AdS black holes were first studied by Hawking and Page
\cite{Hawking:1982dh}, who discovered the Hawking-Page phase transition
between Schwarzschild AdS black holes and thermal AdS space. Later, motivated
by AdS/CFT correspondence \cite{Maldacena:1997re,Gubser:1998bc,Witten:1998qj},
there has been much interest in studying phase structure and transitions of
AdS black holes
\cite{Witten:1998zw,Chamblin:1999tk,Chamblin:1999hg,Caldarelli:1999xj,Cai:2001dz,Kubiznak:2012wp,Wei:2015iwa,Wang:2018xdz,Wei:2019uqg,Wang:2019cax}%
. In light of this, it is of great interest to study spontaneous scalarization
of asymptotically AdS black holes and associated thermodynamic properties in
the EMS models with non-minimal couplings of the scalar and electromagnetic fields.

The remainder of this paper is organized as follows. In section
\ref{sec:Scalarized-Black-Hole}, we introduce the EMS model with a negative
cosmological constant and derive the free energy in a canonical ensemble.
Section \ref{sec:Scalar-Perturbation} is devoted to discussing linear
perturbations in scalar-free and scalarized black hole solutions. In section
\ref{sec:Numeric-Results}, we present our main numerical results, including
the domain of existence, entropic preference, effective potentials for radial
perturbations, and phase structure and transitions in a canonical ensemble. We
summarize our results with a brief discussion in section \ref{sec:Conclusion}.

\section{EMS Model in AdS Space}

\label{sec:Scalarized-Black-Hole}

In this section, we derive the equations of motion, asymptotic behavior, the
Smarr relation and the Helmholtz free energy for asymptotically AdS scalarized
black hole solutions in the EMS model. The action of the EMS model with a
negative cosmological constant is
\begin{equation}
S_{\text{bulk}}=-\frac{1}{16\pi}\int d^{4}x\sqrt{-g}\left[  R-2\Lambda
-2\left(  \partial\phi\right)  ^{2}-f\left(  \phi\right)  F_{\mu\nu}F^{\mu\nu
}\right]  , \label{eq:action}%
\end{equation}
where we take $G=1$ for simplicity throughout this paper. In the action
$\left(  \ref{eq:action}\right)  $, the scalar field $\phi$ is minimally
coupled to the metric $g_{\mu\nu}$ and non-minimally coupled to the gauge
field $A_{\mu}$, $F_{\mu\nu}=\partial_{\mu}A_{\nu}-\partial_{\nu}A_{\mu}$ is
the electromagnetic field strength tensor, $\Lambda=-3/L^{2}$ is the
cosmological constant with the AdS radius $L$, and $f\left(  \phi\right)  $ is
the non-minimal coupling function of the scalar and the gauge fields.

\subsection{Equations of motion}

Varying the action $\left(  \ref{eq:action}\right)  $ with respect to the
metric $g_{\mu\nu}$, the scalar field $\phi$ and the gauge field $A_{\mu}$,
one obtains the equations of motion,
\begin{align}
R_{\mu\nu}-\frac{1}{2}Rg_{\mu\nu}-\frac{3g_{\mu\nu}}{L^{2}}  &  =2T_{\mu\nu
},\nonumber\\
\square\phi-\frac{1}{4}\dot{f}\left(  \phi\right)  F_{\mu\nu}F^{\mu\nu}  &
=0,\label{eq:EOMs}\\
\partial_{\mu}\left(  \sqrt{-g}f\left(  \phi\right)  F^{\mu\nu}\right)   &
=0,\nonumber
\end{align}
where $\dot{f}\left(  \phi\right)  \equiv df\left(  \phi\right)  /d\phi$. The
energy-momentum tensor $T_{\mu\nu}$ in eqn. $\left(  \ref{eq:EOMs}\right)  $
is given by
\begin{equation}
T_{\mu\nu}=\partial_{\mu}\phi\partial_{\nu}\phi-\frac{1}{2}g_{\mu\nu}\left(
\partial\phi\right)  ^{2}+f\left(  \phi\right)  \left(  F_{\mu\rho}F_{\nu
}^{\;\rho}-\frac{1}{4}g_{\mu\nu}F_{\rho\sigma}F^{\rho\sigma}\right)  .
\label{eq:enetgy momentum tensor}%
\end{equation}
In the following, we focus on the spherically symmetric ansatz for the metric,
the electromagnetic field and the scalar field,
\begin{align}
ds^{2}  &  =-N\left(  r\right)  e^{-2\delta\left(  r\right)  }dt^{2}+\frac
{1}{N\left(  r\right)  }dr^{2}+r^{2}\left(  d\theta^{2}+\sin^{2}\theta
d\varphi^{2}\right)  ,\nonumber\\
A_{\mu}dx^{\mu}  &  =V\left(  r\right)  dt\text{ and }\phi=\phi\left(
r\right)  . \label{eq:metric}%
\end{align}
Plugging the above ansatz into the equations of motion $\left(  \ref{eq:EOMs}%
\right)  $ yields
\begin{align}
N^{\prime}\left(  r\right)   &  =\frac{1-N\left(  r\right)  }{r}-\frac{Q^{2}%
}{r^{3}f\left(  \phi\left(  r\right)  \right)  }-rN\left(  r\right)
\phi^{\prime2}\left(  r\right)  +\frac{3r}{L^{2}},\nonumber\\
\left(  r^{2}N\left(  r\right)  \phi^{\prime}\left(  r\right)  \right)
^{\prime}  &  =-\frac{\dot{f}\left(  \phi\left(  r\right)  \right)  Q^{2}%
}{2f^{2}\left(  \phi\right)  r^{2}}-r^{3}N\left(  r\right)  \phi^{\prime
3}\left(  r\right)  ,\nonumber\\
\delta^{\prime}\left(  r\right)   &  =-r\phi^{\prime2}\left(  r\right)
,\label{eq:NLEqs}\\
V^{\prime}\left(  r\right)   &  =-\frac{Q}{r^{2}f\left(  \phi\left(  r\right)
\right)  }e^{-\delta\left(  r\right)  },\nonumber
\end{align}
where primes denote the derivatives with respect to the radial coordinate $r$,
and the integration constant $Q$ can be interpreted as the electric charge of
the black hole solution. For later use, we introduce the Misner-Sharp mass
function $m\left(  r\right)  $ by $N\left(  r\right)  =1-2m\left(  r\right)
/r+r^{2}/L^{2}$.

\subsection{Asymptotic behavior}

To obtain non-trivial hairy black hole solutions of the non-linear ordinary
differential equations $\left(  \ref{eq:NLEqs}\right)  $, one should impose
appropriate boundary conditions at the event horizon and the spatial infinity.
Accordingly, in the vicinity of the event horizon at $r=r_{+}$, we find that
the solutions can be approximated as a power series expansion in terms of
$\left(  r-r_{+}\right)  $,
\begin{align}
m\left(  r\right)   &  =\frac{r_{+}}{2}\left(  1+\frac{r_{+}^{2}}{L^{2}%
}\right)  +m_{1}\left(  r-r_{+}\right)  +\cdots,\delta\left(  r\right)
=\delta_{0}+\delta_{1}\left(  r-r_{+}\right)  +\cdots,\nonumber\\
\phi\left(  r\right)   &  =\phi_{0}+\phi_{1}\left(  r-r_{+}\right)
+\cdots,V\left(  r\right)  =v_{1}\left(  r-r_{+}\right)  +\cdots,
\label{eq:Asols at bounb}%
\end{align}
where
\begin{equation}
m_{1}=\frac{Q^{2}}{2r_{+}^{2}f\left(  \phi_{0}\right)  },\phi_{1}=-\frac
{\dot{f}\left(  \phi_{0}\right)  Q^{2}}{2\left[  f^{2}\left(  \phi_{0}\right)
r_{+}^{3}-f\left(  \phi_{0}\right)  r_{+}Q^{2}+3f^{2}\left(  \phi_{0}\right)
r_{+}^{5}/L^{2}\right]  },\delta_{1}=-r_{+}\phi_{1}^{2},v_{1}=-\frac{Q}%
{r_{+}^{2}f\left(  \phi_{0}\right)  }e^{-\delta_{0}}.
\label{eq:coeff of Asymt. Event}%
\end{equation}
The two essential parameters, $\phi_{0}$ and $\delta_{0}$, can be determined
after matching the asymptotic expansion of the solutions at the spatial
infinity,%
\begin{equation}
m\left(  r\right)  =M-\frac{Q^{2}}{2r}+\cdots,\,\phi\left(  r\right)
=\frac{\phi_{+}}{r^{3}}+\cdots,\delta\left(  r\right)  =\frac{3\phi_{+}^{2}%
}{2r^{6}}+\cdots,\,V\left(  r\right)  =\Phi+\frac{Q}{r}+\cdots,
\label{eq:aymp. at inf}%
\end{equation}
where $f\left(  0\right)  =1$ is assumed, $M$ is identified as the ADM mass,
$\phi_{+}$ can be interpreted as the expectation value of the dual operator in
the conformal boundary theory from AdS/CFT correspondence, and $\Phi$ is the
electrostatic potential with $\Phi=\int_{r_{+}}^{\infty}dre^{-\delta\left(
r\right)  }Q/\left(  r^{2}f\left(  \phi\left(  r\right)  \right)  \right)  $.
Note that the scalar field usually behaves as $\phi\left(  r\right)  \sim
\phi_{-}+\frac{\phi_{+}}{r^{3}}$ in the asymptotic region, and we set
$\phi_{-}=0$ in this paper, corresponding to the absence of the external
source in the conformal boundary theory. Consequently, we can use the shooting
method to solve the non-linear differential equations $\left(  \ref{eq:NLEqs}%
\right)  $ for solutions satisfying the asymptotic expansions at the
boundaries. It is also noteworthy that there is a scaling symmetry among the
physical quantities,%
\begin{equation}
r\rightarrow\lambda r,\,M\rightarrow\lambda M,\,Q\rightarrow\lambda
Q,\,L\rightarrow\lambda L, \label{eq:ss}%
\end{equation}
which allows us to solve eqn. $\left(  \ref{eq:NLEqs}\right)  $ numerically in
terms of redefined dimensionless quantities.

\subsection{Smarr relation}

The Smarr relation \cite{Smarr:1972kt} can be used to test the accuracy of
numerical scalarized black hole solutions, since it associates the black hole
mass with other physical quantities. The Smarr relation can be derived from
computing the Komar integral for a time-like Killing vector $K^{\mu}=\left(
1,0,0,0\right)  $ in a manifold $\mathcal{M}$. Integrating the identity
$\nabla_{\mu}\left(  \nabla_{\nu}K^{\mu}\right)  =K^{\mu}R_{\mu\nu}$ over the
time constant hypersurface $\Sigma$, whose boundary $\partial\Sigma$ consists
of the event horizon $r=r_{+}$ and the spatial infinity $r=+\infty$, one can
use Gauss's law to obtain
\begin{equation}
\int_{\partial\Sigma}dS_{\mu\nu}\nabla^{\mu}K^{\nu}=\int_{\Sigma}dS_{\mu
}K_{\nu}\left(  2T_{\mu\nu}-Tg_{\mu\nu}-\frac{3g_{\mu\nu}}{L^{2}}\right)  ,
\label{eq:Komar Integral}%
\end{equation}
where $dS_{\mu\nu}$ is the surface element on $\partial\Sigma$, and $dS_{\mu}$
is the volume element on $\Sigma$, accordingly. Making use of eqns. $\left(
\ref{eq:enetgy momentum tensor}\right)  $ and $\left(  \ref{eq:NLEqs}\right)
$, we find that the Smarr relation is given by
\begin{equation}
M=\frac{A_{H}T}{2}+Q\Phi-e^{-\delta_{0}}\frac{r_{+}^{3}}{L^{2}}+\int_{r_{+}%
}^{\infty}dre^{-\delta\left(  r\right)  }\delta^{\prime}\left(  r\right)
\frac{r^{3}}{L^{2}}, \label{eq:smarr}%
\end{equation}
where $A_{H}=4\pi r_{+}^{2}$ is the horizon area, and $T=N^{\prime}\left(
r_{+}\right)  e^{-\delta\left(  r_{+}\right)  }/4\pi$ is the Hawking
temperature. For a RNAdS black hole with $\delta\left(  r\right)  =0$, the
Smarr relation $\left(  \ref{eq:smarr}\right)  $ reduces to%
\begin{equation}
M=\frac{A_{H}T}{2}+Q\Phi-\frac{r_{+}^{3}}{L^{2}},
\end{equation}
where the last term is the $PV$ term in the extended phase space of AdS black
holes \cite{Kubiznak:2016qmn}.

\subsection{Free energy}

Given a family of scalarized black holes, it is of interest to compute the
Helmholtz free energy, which can be used to investigate phase structure and
transitions in a canonical ensemble with fixed charge $Q$ and temperature $T$.
The free energy, which is identified as the thermal partition function of
black holes, can be derived via constructing the Euclidean path integral. In
the semiclassical approximation, the partition function is evaluated by
exponentiating the on-shell Euclidean action $S_{\text{on-shell}}^{E}$,
\begin{equation}
Z\sim e^{-S_{\text{on-shell}}^{E}}, \label{eq:partion function}%
\end{equation}
where the on-shell action $S_{\text{on-shell}}^{E}$ is obtained by
substituting the classical solution into the action. However, the on-shell
action $S_{\text{on-shell}}^{E}$ normally diverges in asymptotically AdS
spacetime. One then needs holographic renormalization to remove divergences
appearing in the asymptotic region. There are several methods to regularize
$S_{\text{on-shell}}^{E}$, such as the background-subtraction method
\cite{Chamblin:1999hg} and the Kounterterms method
\cite{Olea:2005gb,Olea:2006vd,Miskovic:2008ck}. Here, we adopt the counterterm
subtraction method \cite{Balasubramanian:1999re,Emparan:1999pm} to regularize
the action by adding a series of boundary terms to the bulk action.

Specifically for the aforementioned bulk action $S_{\text{bulk}}$ in eqn.
$\left(  \ref{eq:action}\right)  $, the regularized action $S_{R}$ is supplied
with three boundary terms%
\begin{equation}
S_{R}=S_{\text{bulk}}+S_{\text{GH}}+S_{\text{ct}}+S_{\text{surf}}\text{,}
\label{eq:Reg. action}%
\end{equation}
where $S_{\text{GH}}$ is the Gibbon-Hawking boundary term to render the
variational problem well-defined, $S_{\text{ct}}$ includes counterterms to
eliminate divergences on asymptotic boundaries, and $S_{\text{surf}}$ is used
to fix the charge rather than the electrostatic potential when the action is
varied \cite{Kim:2016dik,Wang:2018xdz}. The three boundary terms are given by
\begin{align}
S_{\text{GH}}  &  =-\frac{1}{8\pi}\int d^{3}x\sqrt{-\gamma}\Theta,\nonumber\\
S_{\text{ct}}  &  =\frac{1}{8\pi}\int d^{3}x\sqrt{-\gamma}\left(  \frac{2}%
{L}+\frac{L}{2}R_{3}\right)  ,\label{eq:boundary terms}\\
S_{\text{surf}}  &  =-\frac{1}{4\pi}\int d^{3}x\sqrt{-\gamma}f\left(
\phi\right)  F^{\mu\nu}n_{\mu}A_{\nu},\nonumber
\end{align}
where the integrals are performed on the hypersurface at the spatial infinity,
$\gamma$ is the determinant of the induced metric on the hypersurface,
$\Theta$ is the trace of the extrinsic curvature, $R_{3}$ is the scalar
curvature of the induced metric $\gamma$, and $n_{\mu}$ is the unit vector
normal to the hypersurface. Using the equations of motion $\left(
\ref{eq:EOMs}\right)  $ and the asymptotic expansion $\left(
\ref{eq:aymp. at inf}\right)  $, we obtain the on-shell Euclidean version of
$S_{\text{bulk}}$, $S_{\text{GH}}$, $S_{\text{ct}}$ and $S_{\text{surf}}$,
\begin{align}
S_{\text{bulk, on-shell}}^{E}  &  =\frac{1}{T}\left(  \left.  \frac
{e^{-\delta\left(  r\right)  }r^{2}N^{\prime}\left(  r\right)  -2e^{-\delta
\left(  r\right)  }r^{2}N\left(  r\right)  \delta^{\prime}\left(  r\right)
}{4}\right\vert _{r=+\infty}-TS-Q\Phi\right)  ,\nonumber\\
S_{\text{GH, on-shell}}^{E}  &  =\left.  -\frac{1}{T}\left[  \frac
{e^{-\delta\left(  r\right)  }r^{2}N^{\prime}\left(  r\right)  -2e^{-\delta
\left(  r\right)  }r^{2}\delta^{\prime}\left(  r\right)  N\left(  r\right)
}{4}+e^{-\delta\left(  r\right)  }\left(  r-2M+\frac{r^{3}}{L^{2}}\right)
\right]  \right\vert _{r=+\infty},\nonumber\\
S_{\text{ct, on-shell}}^{E}  &  =\left.  \frac{e^{-\delta\left(  r\right)  }%
}{T}\left(  \frac{r^{3}}{L^{2}}+r-M\right)  \right\vert _{r=+\infty},\\
S_{\text{surf, on-shell}}^{E}  &  =\frac{Q\Phi}{T},\nonumber
\end{align}
where $S=\pi r_{+}^{2}$ is the entropy of the black hole. Consequently, the
regularized on-shell Euclidean action for the black hole solution $\left(
\ref{eq:metric}\right)  $ is
\begin{equation}
S_{\text{on-shell}}^{E}=S_{\text{bulk, on-shell}}^{E}+S_{\text{GH, on-shell}%
}^{E}+S_{\text{ct, on-shell}}^{E}+S_{\text{surf, on-shell}}^{E}=\frac{M-TS}%
{T}.
\end{equation}
The Helmholtz free energy $F$ is related to the Euclidean action
$S_{\text{on-shell}}^{E}$ via%
\begin{equation}
F=-T\ln Z=TS_{\text{on-shell}}^{E},
\end{equation}
which gives%
\begin{equation}
F=M-TS. \label{eq:freenergy}%
\end{equation}

\section{Perturbations around Black Hole Solution}

\label{sec:Scalar-Perturbation}

In this section, we investigate linear perturbations around black hole
solutions, which can help us understand the stability of the solutions.

\subsection{Scalar perturbation around RNAdS black holes}

We first examine a scalar perturbation $\delta\phi$ in a RNAdS black hole
background. Note that if $\dot{f}\left(  0\right)  =0$ is imposed, a RNAdS
black hole, which is described by
\begin{equation}
N\left(  r\right)  =1-\frac{2M}{r}+\frac{Q^{2}}{r^{2}}+\frac{r^{2}}{L^{2}%
},A=\frac{Q}{r}dt,\delta\left(  r\right)  =0,\phi\left(  r\right)  =0,
\label{eq:RNAdS solution}%
\end{equation}
is manifestly a solution of the equations of motion $\left(  \ref{eq:EOMs}%
\right)  $. In this scalar-free solution background, we can linearize the
scalar equation in eqn. $\left(  \ref{eq:EOMs}\right)  $ with a scalar
perturbation $\delta\phi$,
\begin{equation}
\left(  \square-\mu_{eff}^{2}\right)  \delta\phi=0,
\label{eq:perturbative equation}%
\end{equation}
where $\mu_{eff}^{2}=-\ddot{f}\left(  0\right)  Q^{2}/\left(  2r^{4}\right)
$. In a $\left(  3+1\right)  $-dimensional asymptotically AdS spacetime of AdS
radius $L$, a scalar field can cause a tachyonic instability only if its
mass-squared is less than the so-called Breitenlohner-Freedman (BF) bound
$\mu_{BF}^{2}=-9/\left(  4L^{2}\right)  $ \cite{Breitenlohner:1982jf}. For the
scalar perturbation $\delta\phi$, one always has $\mu_{eff}^{2}>\mu_{BF}^{2}$
for large enough $r$, and hence, asymptotically, the RNAdS black hole is
stable against the formation of the scalar field, which guarantees that
scalarized black holes induced by the tachyonic instability of the scalar
field are asymptotically AdS. However, if $\mu_{eff}^{2}<\mu_{BF}^{2}$ in some
region (e.g., near the event horizon), a RNAdS black hole may evolve to a
scalarized black hole under a scalar perturbation. Note that if $\ddot
{f}\left(  0\right)  <0$, one always has $\mu_{eff}^{2}>\mu_{BF}^{2}$, and
hence a tachyonic instability can not occur.

To study how a scalarized black hole solution bifurcates from a scalar-free
black hole solution, we calculate zero modes of the scalar perturbation in the
scalar-free black hole background. For simplicity, the scalar perturbation
$\delta\phi$ is written as the decomposition with spherical harmonics
functions,
\begin{equation}
\delta\phi=\sum_{l,m}Y_{lm}\left(  \theta,\phi\right)  U_{l}\left(  r\right)
.
\end{equation}
With this decomposition, the scalar equation $\left(
\ref{eq:perturbative equation}\right)  $ then reduces to
\begin{equation}
\frac{1}{r^{2}}\frac{d}{dr}\left(  r^{2}N\left(  r\right)  \frac{dU_{l}\left(
r\right)  }{dr}\right)  -\left[  \frac{l\left(  l+1\right)  }{r^{2}}+\mu
_{eff}^{2}\right]  U_{l}\left(  r\right)  =0. \label{eq:seqn}%
\end{equation}
Given the fixed values of $l$ and $\ddot{f}\left(  0\right)  $, requiring that
the radial field $U_{l}\left(  r\right)  $ is regular at the event horizon and
vanishes at the spatial infinity selects a family of discrete black hole
solutions, which can be labelled by a non-negative integer node number $n$. In
this paper, we focus on the $l=0=n$ fundamental mode since it gives the
smallest $q$ of the black hole solutions \cite{Herdeiro:2018wub}. Due to the
tachyonic instability, scalarized RNAdS black holes may emerge from these zero
modes, which composes bifurcation lines in the domain of existence for the
scalarized black holes.

\subsection{Time-dependent perturbation around scalarized black holes}

To investigate perturbative stability of the scalarized black hole solution
$\left(  \ref{eq:metric}\right)  $, we then consider spherically symmetric and
time-dependent linear perturbations. Specifically including the perturbations,
the metric ansatz is written as \cite{Fernandes:2019rez}%
\begin{align}
ds^{2}  &  =-\tilde{N}\left(  r,t\right)  e^{-2\tilde{\delta}\left(
r,t\right)  }dt^{2}+\frac{dr^{2}}{\tilde{N}\left(  r,t\right)  }+r^{2}\left(
d\theta^{2}+\sin^{2}\theta d\varphi^{2}\right)  ,\nonumber\\
\tilde{N}\left(  r,t\right)   &  =N\left(  r\right)  +\epsilon\tilde{N}%
_{1}\left(  r\right)  e^{-i\Omega t},\tilde{\delta}\left(  r,t\right)
=\delta\left(  r\right)  +\epsilon\tilde{\delta}_{1}\left(  r\right)
e^{-i\Omega t}, \label{eq:time-dependent metric}%
\end{align}
where the time dependence of the perturbations is assumed to be Fourier modes
with frequency $\Omega$. Similarly, the ansatzes of the scalar and
electromagnetic fields are given by
\begin{equation}
\tilde{\phi}\left(  r,t\right)  =\phi\left(  r\right)  +\epsilon\phi
_{1}\left(  r\right)  e^{-i\Omega t}\;\text{and}\;\tilde{V}\left(  r,t\right)
=V\left(  r\right)  +\epsilon V_{1}\left(  r\right)  e^{-i\Omega t},
\label{eq:time-dependent fields}%
\end{equation}
respectively. Solving eqn. $\left(  \ref{eq:EOMs}\right)  $ with the ansatzes
$\left(  \ref{eq:time-dependent metric}\right)  $ and $\left(
\ref{eq:time-dependent fields}\right)  $, we can extract a Schrodinger-like
equation for the perturbative scalar field $\phi_{1}\left(  r\right)  $,
\begin{equation}
-\frac{d^{2}\Psi\left(  r\right)  }{dr^{\ast2}}+U_{\Omega}\Psi\left(
r\right)  =\Omega^{2}\Psi\left(  r\right)  ,
\end{equation}
where $\Psi\left(  r\right)  \equiv r\phi_{1}\left(  r\right)  $, and the
tortoise coordinate $r^{\ast}$ is defined by $dr^{\ast}/dr\equiv
e^{\delta\left(  r\right)  }N^{-1}\left(  r\right)  $. Here, the effective
potential $U_{\Omega}$ is given by
\begin{equation}
U_{\Omega}=\frac{e^{-2\delta}N}{r^{2}}\left\{  1-N-2r^{2}\phi^{\prime2}%
-\frac{Q^{2}}{r^{2}f\left(  \phi\right)  }\left(  1-2r^{2}\phi^{\prime2}%
+\frac{\ddot{f}\left(  \phi\right)  }{2f\left(  \phi\right)  }+2r\phi^{\prime
}\frac{\dot{f}\left(  \phi\right)  }{f\left(  \phi\right)  }-\frac{\dot{f}%
^{2}\left(  \phi\right)  }{f^{2}\left(  \phi\right)  }\right)  +\frac
{3r^{2}\left(  1-2r^{2}\phi^{\prime2}\right)  }{L^{2}}\right\}  .
\label{eq:effe.  potential}%
\end{equation}
One can show that the effective potential $U_{\Omega}$ vanishes at the event
horizon, whereas approaches positive infinity at the spatial infinity. From
quantum mechanics, the existence of an unstable mode with $\Omega^{2}<0$
requires the presence of $U_{\Omega}<0$ in some regions. Nevertheless, a
positive definite $U_{\Omega}$ ensures that scalarized black hole solutions
are stable against the spherically symmetric perturbations. It is noteworthy
that the appearance of a negative region of $U_{\Omega}$ cannot sufficiently
guarantee the presence of an instability \cite{Zou:2019bpt}. One can utilize
other techniques, like the $S$-deformation method \cite{Kimura:2017uor}, to
further discuss the stability of these solutions.

\section{Numerical Results}

\label{sec:Numeric-Results}

In this section, we first study various properties of scalarized RNAdS black
hole solutions and then investigate their phase structure and transitions in a
canonical ensemble. After the non-linear differential equations $\left(
\ref{eq:NLEqs}\right)  $ are expressed in terms of a new dimensionless
coordinate%
\begin{equation}
x=1-\frac{r_{+}}{r}\text{ with }0\leq x\leq1\text{,}%
\end{equation}
they are numerically solved for scalarized black hole solutions using the
NDSolve function in Wolfram Mathematica. In the remainder of this paper, we
focus on the coupling function $f\left(  \phi\right)  =e^{\alpha\phi^{2}}$
with $\alpha>0$. For this coupling function, one has $f\left(  0\right)  =1$
and $\dot{f}\left(  0\right)  =0$, which ensures that RNAdS black holes are
solutions of the EMS model, and $\ddot{f}\left(  0\right)  >0$, which may
trigger a tachyonic instability of the scalar field to induce scalarized black
hole solutions. For later use, we define reduced quantities,%
\begin{equation}
q=\frac{Q}{M},a_{H}=\frac{A_{H}}{16\pi M^{2}},\tilde{T}=TL,\,\tilde{F}%
=\frac{F}{L},\,\tilde{r}_{+}=\frac{r_{+}}{L},\,\tilde{Q}=\frac{Q}{L}%
,\,\tilde{M}=\frac{M}{L},
\end{equation}
which are dimensionless and invariant under the scaling symmetry $\left(
\ref{eq:ss}\right)  $. To test the accuracy of our numerical method, we use
the Smarr relation $\left(  \ref{eq:smarr}\right)  $ and find that the
numerical error can be maintained around the order of $10^{-6}$.

\subsection{Scalarized black holes}

\begin{figure}[ptb]
\includegraphics[scale=0.85]{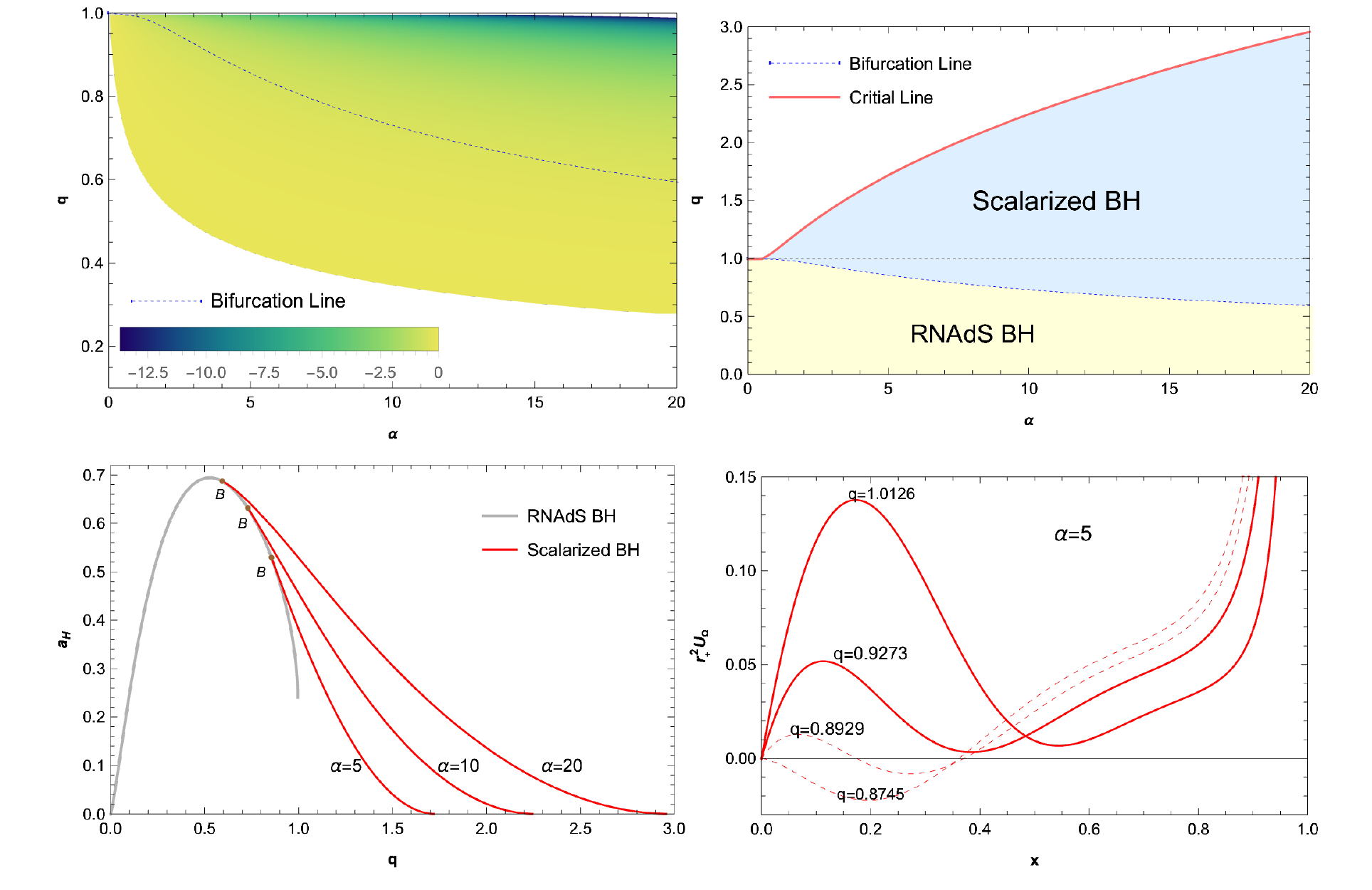}\caption{Plots of tachyonically unstable
region for RNAdS black holes, and the domain of existence, entropic preference
and effective potentials for scalarized black holes. Here, we take $Q/L=0.1$.
\textbf{Upper Left:} Density plot of $\left(  \mu_{eff}^{2}-\mu_{BF}%
^{2}\right)  Q^{2}$ evaluated at the event horizon $r=r_{+}$ as a function of
$q$ and $\alpha$ for the scalar perturbation in RNAdS black holes. We only
display the tachyonically unstable region, where $\mu_{eff}^{2}<\mu_{BF}^{2}$,
in the $\alpha\text{-}q$ plane. The closer RNAdS black holes are to the
extremal limit, the more unstable the scalar field becomes. The blue dashed
line represents the bifurcation line, where tachyonic instabilities are strong
enough to induce scalarized black holes. \textbf{Upper Right: }Domain of
existence for scalarized RNAdS black holes in the $\alpha\text{-}q$ plane,
which is highlighted by the light blue region and bounded by the critical and
bifurcation lines. The critical line is depicted by a red solid line, on which
the reduced horizon area $a_{H}$ vanishes. The horizontal dashed gray line
denotes extremal RNAdS black holes, above which RNAdS black hole solutions do
not exist. \textbf{Lower Left: }Reduced horizon area $a_{H}$ against $q$ for
RNAdS and scalarized black holes. The scalarized black hole solutions are
always entropically preferred, which means that they are globally stable in a
micro-canonical ensemble. \textbf{Lower Right:} Effective potentials of
scalarized black holes with $\alpha=5$ for several values of $q$. Solid red
lines denote positive definite effective potentials between the event horizon
and the spatial infinity, while dashed red lines represent those possessing
negative regions. When $q$ is large enough, the scalarized black hole
solutions are stable against radial perturbations.}%
\label{Fig1}%
\end{figure}

Here, we present the numerical results, e.g., the domain of existence,
entropic preference and effective potentials, for scalarized black hole
solutions, which are dynamically induced from RNAdS black holes. Without loss
of generality, we focus on $\tilde{Q}=Q/L=0.1$ in this subsection. For a RNAdS
black hole, a tachyonic instability of the scalar field occurs if $\mu
_{eff}^{2}<\mu_{BF}^{2}$ somewhere in the spacetime. Since the minimum value
of $\mu_{eff}^{2}$ occurs at the event horizon $r=r_{+}$, we only need to
check $\mu_{eff}^{2}<\mu_{BF}^{2}$ at $r=r_{+}$. The region in the $\left(
\alpha,q\right)  $ parameter space of RNAdS black holes where $\mu_{eff}%
^{2}<\mu_{BF}^{2}$ at $r=r_{+}$ is plotted in the upper left panel of Fig.
\ref{Fig1}. The distribution of values of $(\mu_{eff}^{2}-\mu_{BF}^{2}%
)Q^{2}|_{r=r_{+}}$ is also displayed, which shows that the tachyonic
instability region becomes larger as $\alpha$ increases, and the scalar field
suffers from a strong tachyonic instability when black holes are
near-extremal. The bifurcation line is composed of the $l=0=n$ zero modes of
eqn. $\left(  \ref{eq:seqn}\right)  $ for the scalar perturbation in RNAdS
black holes, and represented by blue dashed lines in Fig. \ref{Fig1}. When the
tachyonic instability is strong enough, the scalar perturbation can lead to
scalarized black holes with non-trivial scalar fields above the bifurcation
line. In the upper right panel of Fig. \ref{Fig1}, we display the domain of
existence for scalarized black holes, which is exhibited by a light blue
region. The domain of existence is bounded by the bifurcation and critical
lines, and resembles that of RN scalarized black holes \cite{Herdeiro:2018wub}%
. On the critical line, the mass and the charge of scalarized solutions remain
finite, whereas its horizon radius vanishes. On the other hand, the
mass-to-charge ratio $q$ of RNAdS black holes reaches the maximum in the
extremal limit, which is shown by a horizontal dashed gray line. Moreover,
there is a certain region bounded by the extremal and bifurcation lines, where
scalarized and RNAdS black holes coexist.

The reduced area $a_{H}$ as a function of reduced charge $q$ is plotted for
RNAdS and scalarized black holes in the lower left panel of FIG. \ref{Fig1},
which demonstrates that scalarized black holes emerge from RNAdS black holes
at the bifurcation points, marked by $B$, and eventually terminate on the
critical line with zero $a_{H}$. For a multiphase system in a micro-canonical
ensemble with conserved energy, the phase of maximum entropy is globally
stable and will be present at equilibrium. Therefore, in the scalarized and
RNAdS black holes coexisting region, our numerical results show that
scalarized solutions are entropically preferred over RNAdS black hole
solutions, and hence are the globally stable phase in the micro-canonical ensemble.

To study the stability of scalarized solutions, the effective potentials
$U_{\Omega}$ of scalarized solutions with $\alpha=5$ are plotted for several
values of $q$ in the lower right panel of FIG. \ref{Fig1}, where solid and
dashed colored lines correspond to potentials with and without negative
regions, respectively. The scalarized solutions have positive effective
potentials for a large enough value of $q$, and thus are free of radial
instabilities. However, as $q$ decreases towards the bifurcation line, there
appear negative regions in the effective potentials, which means that radial
instabilities cannot be excluded near the bifurcation line.

\subsection{Phase structure in a canonical ensemble}

\begin{figure}[ptb]
\includegraphics[scale=0.65]{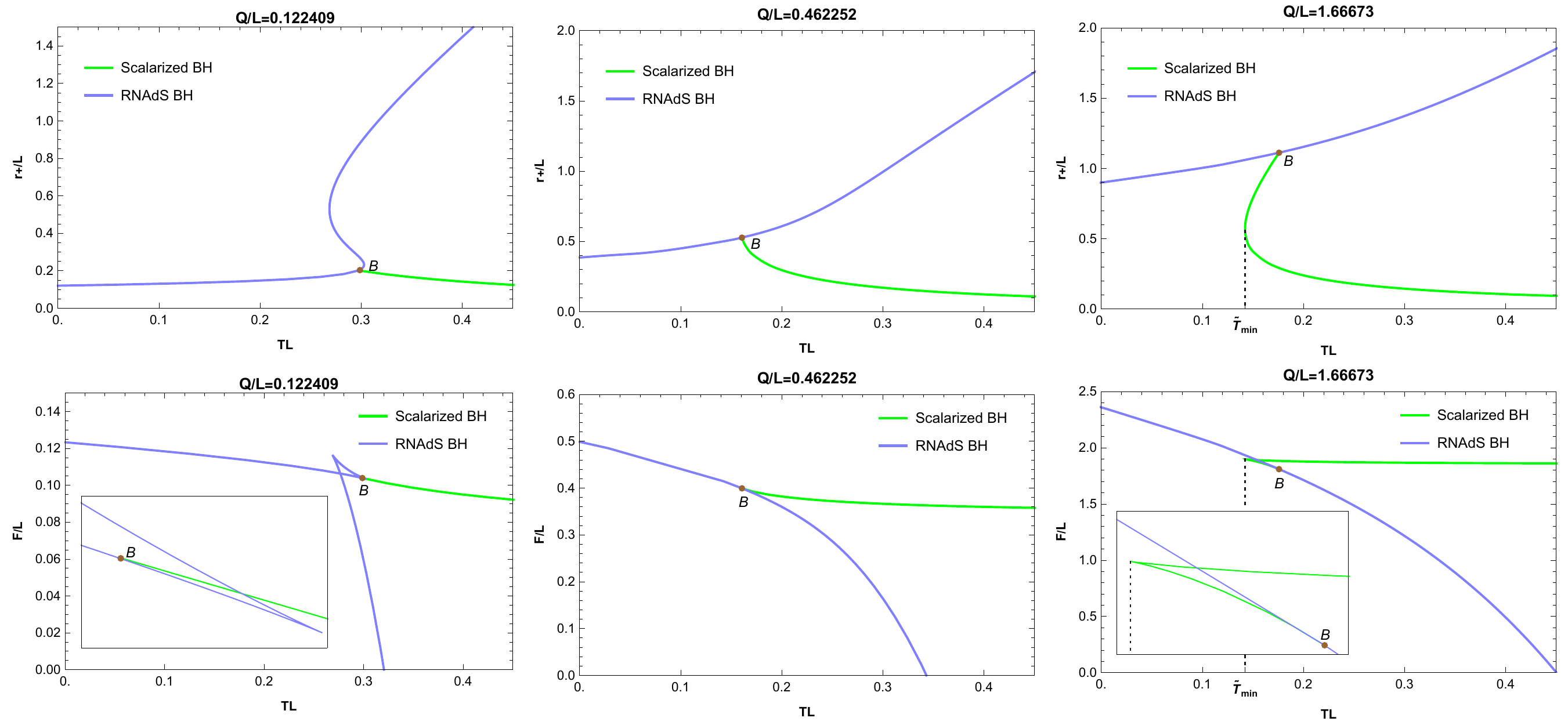}\caption{Plots of the reduced horizon radius
$\tilde{r}_{+}$ (the upper row) and the reduced free energy $\tilde{F}$ (the
lower row) versus the reduced temperature $\tilde{T}$ for RNAdS (blue lines)
and scalarized (green lines) black holes with several fixed values of the
reduced charge $\tilde{Q}$. Here, we focus on $\alpha=5$. Bifurcation points
are labeled by $B$. When $\tilde{r}_{+}(\tilde{T})$ is multivalued, black hole
solutions have more than one branch of different horizon radii in a canonical
ensemble with fixed $\tilde{T}$ and $\tilde{Q}$. \textbf{Left column}: There
is a band of temperatures where three branches of RNAdS black hole solutions
coexist, and a first-order phase transition occurs between the large RNAdS BH
phase (i.e., the branch with the largest horizon radius) and the small RNAdS
BH phase (i.e., the branch with the smallest horizon radius). Scalarized black
holes emerge from the bifurcation point. Nevertheless, they are not globally
preferred since they always have a higher free energy than RNAdS black holes.
\textbf{Center column}: RNAdS black hole solutions have only one branch, whose
free energy is smaller than that of scalarized black holes. There is no phase
transition. \textbf{Right column}: RNAdS black hole solutions have only one
branch, whereas scalarized black hole solutions have two branches of different
sizes. As $\tilde{T}$ increases, the globally stable phase jumps from RNAdS
black holes to scalarized black holes (the branch with a larger horizon
radius), corresponding to a zeroth-order phase transition at $\tilde{T}_{\min
}$. Further increasing $\tilde{T}$, there would be a second-order phase
transition returning to RNAdS black holes at the bifurcation point $B$. Here,
we observe a RNAdS BH/scalarized BH/RNAdS BH reentrant phase transition.}%
\label{Fig2}%
\end{figure}

In this subsection, we consider phase structure and transitions of scalarized
and RNAdS black holes in a canonical ensemble maintained at a given
temperature of $T$ and a given charge of $Q$. In a canonical ensemble, the
globally stable phase of a multiphase system, which exists at equilibrium, has
the lowest possible Helmholtz free energy $F$, which can be computed via eqn.
$\left(  \ref{eq:freenergy}\right)  $. The rich phase structure of black holes
usually comes from expressing the horizon radius $r_{+}$ as a function of
temperature $T$. If the function $r_{+}\left(  T\right)  $ is multivalued,
there will be more than one black hole phase, corresponding to different
branches of $r_{+}\left(  T\right)  $.

To illustrate phase structure and transitions, we plot the reduced horizon
radius $\tilde{r}_{+}$ and the free energy $\tilde{F}$ as functions of reduced
temperature $\tilde{T}$ for scalarized and RNAdS black holes with three
representative values of $\tilde{Q}$ in Fig. \ref{Fig2}, where we have
$\alpha=5$. In the left column of Fig. \ref{Fig2} with a small $\tilde{Q}$,
the upper panel shows that three branches of the RNAdS black hole solution
coexist in some range of $\tilde{T}$, and are dubbed as large, intermediate
and small RNAdS BHs, respectively, based on their values of horizon radius. At
a high (low) enough temperature, only the large (small) RNAdS BH phase exists.
On the other hand, there is only one phase for the scalarized black hole
solution, which bifurcates from the RNAdS black hole solution at the
bifurcation point $B$, and does not exist at a low temperature. The reduced
free energy $\tilde{F}$ is plotted against $\tilde{T}$ for these four phases
in the lower panel, which shows that the scalarized black hole can not be the
globally stable phase since there always exists some RNAdS black hole phase of
a lower free energy at a given $\tilde{T}$. In the coexisting region of the
RNAdS black hole phases, a first-order phase transition between large and
small RNAdS BHs occurs at some point, where their free energies intersect each other.

The coexisting region of the RNAdS black hole phases shrinks as $\tilde{Q}$
increases until reaching a critical point, where a second-order phase
transition occurs between large and small RNAdS BHs. Beyond the critical
point, large RNAdS BH is indistinguishable from small RNAdS BH, hence RNAdS
black hole solutions have a single phase. In the center column of Fig.
\ref{Fig2}, we depict $\tilde{r}_{+}(\tilde{T})$ and $\tilde{F}(\tilde{T})$
for the case with a value of $\tilde{Q}$ greater than the critical value. The
upper panel shows that both RNAdS and scalarized black holes have a single
phase. Moreover, as displayed in the lower panel, the RNAdS black hole always
has a smaller free energy than the scalarized black hole, and hence is the
globally stable phase. Consequently, there is no phase transition in this case.

However when $\tilde{Q}$ is large enough, phase structure of scalarized and
RNAdS black holes becomes much richer. For example, $\tilde{r}_{+}(\tilde{T})$
and $\tilde{F}(\tilde{T})$ are plotted for scalarized and RNAdS black holes
with a large enough $\tilde{Q}$ in the right column of Fig. \ref{Fig2}. It
shows in the upper panel that the RNAdS black hole solution possesses a single
phase, whereas the scalarized solution can have two phases at some given
$\tilde{T}$, namely large scalarized BH (i.e., the one with a larger horizon
radius) and small scalarized BH (i.e., the one with a smaller horizon radius).
In fact, the scalarized black hole solution has a minimum temperature
$\tilde{T}_{\min}$, and large scalarized BH coexists with small scalarized BH
between $\tilde{T}=\tilde{T}_{\min}$\ and the bifurcation point $B$, where
large scalarized BH and the RNAdS black hole merge. The lower panel exhibits
the free energy as a function of $\tilde{T}$ for the three phases. If we start
increasing the temperature from $\tilde{T}=0$, the system follows the blue
line of the RNAdS black hole until $\tilde{T}=\tilde{T}_{\min}$, where the
free energy has a discontinuity at its global minimum. Further increasing
$\tilde{T}$, the inset shows that the system jumps to the lower green line of
large scalarized BH, which corresponds to a zeroth-order phase transition
between scalarized and RNAdS black holes. As $\tilde{T}$ continues to
increase, the system follows the lower green line until it joins the blue line
at the bifurcation point $B$, which corresponds to a second-order phase
transition between scalarized and RNAdS black holes. Note that since
scalarized and RNAdS black holes have the same entropy at the bifurcation
point, a phase transition between them at the bifurcation point is
second-order. In short, a RNAdS BH/scalarized BH/RNAdS BH reentrant phase
transition is observed as $\tilde{T}$ increases.

In addition, it is interesting to consider the local thermodynamic stability
of black hole phases against thermal fluctuations. In a canonical ensemble,
the quantity of particular interest is the specific heat at constant charge,%
\begin{equation}
C_{Q}=T\left(  \frac{\partial S}{\partial T}\right)  _{Q}=2L^{2}\pi\tilde
{r}_{+}\tilde{T}\frac{\partial\tilde{r}_{+}}{\partial\tilde{T}}. \label{eq:CQ}%
\end{equation}
Since the entropy is proportional to the size of the black hole, a positive
specific heat means that black holes radiate less when they are smaller. Thus,
the thermal stability of a phase follows from $C_{Q}>0$ (or equivalently,
$\partial\tilde{r}_{+}/\partial\tilde{T}>0$). From the upper row of Fig.
\ref{Fig2}, we notice that large and small RNAdS BHs in the left column, RNAdS
black hole in the center column, and large scalarized BH and RNAdS black hole
in the right column all have $\partial\tilde{r}_{+}/\partial\tilde{T}>0$. In
consequence, the globally stable phases of scalarized and RNAdS black holes
possess a positive $C_{Q}$, and are thermally stable.

\begin{figure}[ptb]
\begin{centering}
\includegraphics[scale=0.7]{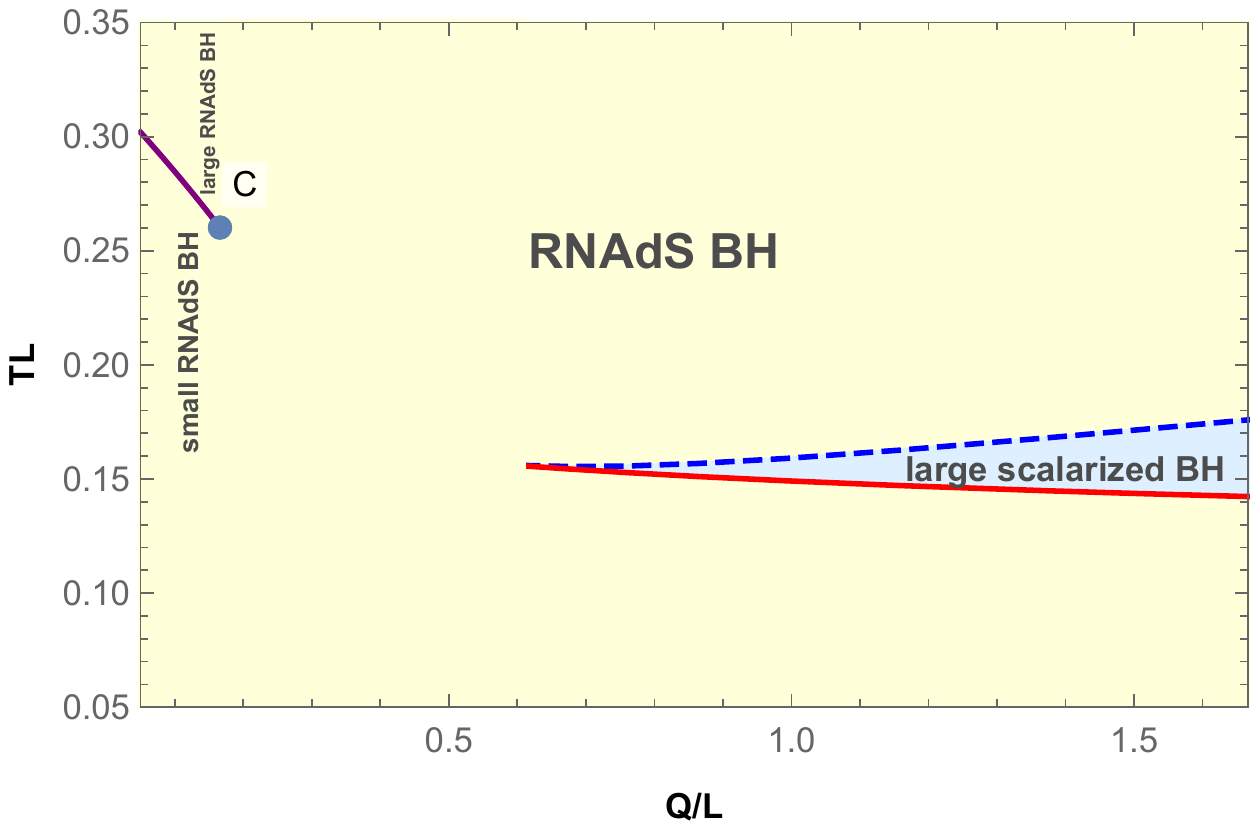}
\par\end{centering}
\caption{Phase diagram of RNAdS and scalarized black holes in a canonical
ensemble with fixed temperature $T$ and charge $Q$. Here, we take $\alpha=5$.
The phase diagram exhibits the globally stable phases, which have the lowest
free energy, and the phase transitions between them. The light yellow and blue
regions correspond to RNAdS and scalarized black holes, respectively. A
first-order phase transition line (the purple line) separates the large RNAdS
BH phase, which is above the line, and the small RNAdS BH phase, which is
below the line. The first-order phase transition line terminates at the
critical point, labelled by $C$. The scalarized black hole phase is delimited
by a zeroth-order phase transition line (the red line) and a second-order one
(the blue dashed line), which coincides with the bifurcation line. }%
\label{Fig3}%
\end{figure}

To better illustrate the globally stable phases of lowest free energy and the
associated phase transitions, we display the phase diagram of scalarized and
RNAdS black holes in the $\tilde{Q}$-$\tilde{T}$ plane in Fig. \ref{Fig3},
where $\alpha=5$. There is a first-order phase transition line (the purple
line) separating large and small RNAdS BHs for small $\tilde{Q}$, which
terminates at the critical point. This first-order phase transition is quite
similar to the liquid/gas phase transition. When $\tilde{Q}$ is large enough,
large scalarized BH (the light blue region) appears, and is bounded by
zeroth-order (the red line) and second-order (the blue dashed line) phase
transition lines.

\section{Discussion and Conclusions}

\label{sec:Conclusion}

In this paper, we investigated spontaneous scalarization of asymptotically AdS
charged black holes in the EMS model, and studied phase structure of
scalarized and RNAdS black holes in a canonical ensemble. We focused on a
non-minimal coupling function $f\left(  \phi\right)  =e^{\alpha\phi^{2}}$,
which leads to spontaneous scalarization due to the tachyonic instability of
the scalar field near the event horizon. In practice, scalarized black holes
can be induced from RNAdS black holes on the bifurcation line, which consists
of zero modes of the scalar perturbation in RNAdS black holes. In the
$\alpha\text{-}q$ plane with a fixed $\tilde{Q}$, the domain of existence for
scalarized RNAdS black holes is bounded by the bifurcation and critical lines,
which resembles that of scalarized RN black holes very closely
\cite{Herdeiro:2018wub}. In a micro-canonical ensemble, we found that
scalarized black hole solutions are always entropically preferred over RNAdS
black holes, and hence the globally stable phase.

On the other hand, the system has much richer phase structure in a canonical
ensemble. After the Helmholtz free energy of the EMS model was computed, we
obtained the phase structure of scalarized and RNAdS black holes. In the small
$Q$ regime, scalarized black holes never globally minimize the free energy,
and the corresponding phase diagram resembles that of the liquid/gas system
closely. Nevertheless in the large $Q$ regime, scalarized black holes can be
the globally stable phase in some parameter region. As the temperature
increases at a given charge, the system undergoes a RNAdS BH/scalarized
BH/RNAdS BH reentrant phase transition, which consists of zeroth-order and
second-order phase transitions.

The phenomenon of reentrant phase transition was first observed in a
nicotine/water mixture, and later discovered in the context of black hole
thermodynamics, e.g., Born-Infeld-AdS black holes
\cite{Gunasekaran:2012dq,Wang:2018xdz}, higher dimensional singly spinning
Kerr-AdS black holes \cite{Altamirano:2013ane}, AdS black holes in Lovelock
gravity \cite{Frassino:2014pha}, AdS black holes in dRGT massive gravity
\cite{Zou:2016sab}, hairy AdS black holes \cite{Hennigar:2015wxa}. For these
black holes, reentrant phase transitions were found to include zeroth-order
and first-order phase transitions. In this paper, we present an example of a
reentrant phase transition for black holes, which is composed of zeroth-order
and second-order phase transitions. Furthermore, the second-order phase
transition between scalarized and RNAdS black holes is of great interest since
this implies that our results may provide an interesting\ model of holographic
superconductors. We leave this for future work.

AdS black holes can also been studied in the context of extended phase space
thermodynamics, where the cosmological constant is interpreted as a
thermodynamic pressure $P\equiv6/L^{2}$ \cite{Dolan:2011xt,Kubiznak:2012wp}.
In terms of $P$, the reduced quantities are expressed as
\begin{equation}
\tilde{T}=T\sqrt{6/P},\tilde{F}=F\sqrt{P/6},\tilde{r}_{+}=r_{+}\sqrt
{P/6},\tilde{Q}=Q\sqrt{P/6},\tilde{M}=M\sqrt{P/6}. \label{eq:ssP}%
\end{equation}
Note that $M$ and $F$ are identified as the enthalpy and the Gibbs free
energy, respectively, in extended phase space. With eqn. $\left(
\ref{eq:ssP}\right)  $, our results can be generalized to extended phase space thermodynamics.

\begin{acknowledgments}
We are grateful to Qingyu Gan and Feiyu Yao for useful discussions and
valuable comments. This work is supported in part by NSFC (Grant No. 11875196,
11375121, 11947225 and 11005016).
\end{acknowledgments}

\bibliographystyle{unsrturl}
\bibliography{ref}

\end{document}